\documentclass[aps,prl,amsmath,amssymb,superscriptaddress,twocolumn]{revtex4-2}
\usepackage{float}
\usepackage{xr}
\usepackage{soul}
\usepackage{graphicx} 
\usepackage{hyperref}

\begin{document}

\title{Topological origin of peak splitting in the structure factor of liquid water}

\author{Zoé Faure Beaulieu}
\affiliation{Inorganic Chemistry Laboratory, Department of Chemistry, University of Oxford, Oxford, UK}
\author{Volker L. Deringer}
\email{volker.deringer@chem.ox.ac.uk}
\affiliation{Inorganic Chemistry Laboratory, Department of Chemistry, University of Oxford, Oxford, UK}
\author{Fausto Martelli}
\email{fausto.martelli@cnr.it}
\affiliation{CNR-Istituto dei Sistemi Complessi, Piazzale Aldo Moro 5, Rome, 00185, Italy}
\affiliation{Dipartimento di Fisica, Universit\`a degli Studi di Roma La Sapienza, Piazzale Aldo Moro 5, Rome, 00185, Italy}



\keywords{water $|$ structure factor $|$ ring statistics $|$ machine-learning potential}

\begin{abstract}
The splitting of the principal peak in the structure factor of liquid water is commonly interpreted as evidence of a competition between two distinct local environments. Here, we show that this peak splitting arises from medium-range topological features of the hydrogen-bond network. Using atomistic simulations, we systematically decompose the structure factor into contributions from hydrogen-bonded rings of different sizes. We find that 5–8-membered rings, which dominate the network topology of liquid water at low temperatures, can directly explain the experimentally observed bimodal scattering signal. Among these, 5-membered rings are particularly persistent, maintaining distinct structural signatures even above room temperature. Our findings establish a direct link between the network topology of liquid water and experimentally accessible diffraction features, clarifying the microscopic basis of water’s behavior and suggesting a broader conceptual framework for interpreting the anomalies in tetrahedral network liquids and glasses.
\end{abstract}


\maketitle

{\em Introduction}---The most direct experimental probe for the structure of a liquid is the X-ray or neutron structure factor, $S(q)$. For simple liquids modelled by Lennard-Jones potentials, the first peak in the structure factor appears at the wavenumber corresponding to the average interparticle distance. However, this trend does not hold for liquids with local tetrahedral ordering, such as water~\cite{shi2020direct}, silicon~\cite{shi2019distinct}, and silica~\cite{Wright-94-11,gaskell1996medium,Mei-08-10}: here, the main peak in the $S(q)$ is often observed at a lower wavenumber. The latter feature is referred to as the first sharp diffraction peak (FSDP), and its origin had long remained unclear~\cite{elliott1991medium,elliott1991origin,massobrio2001origin,phillips1981topology}.
Shi and Tanaka have shown that the FSDP in covalent liquids (including water) and glasses is a combination of two overlapping peaks: one arising from the density-wave characteristic of local structures lacking tetrahedral symmetry, and the other arising from the density-wave characteristic of tetrahedral local environments~\cite{shi2019distinct,shi2020direct}. This argument supports the view that water behaves according to a two-state model, i.e., as a mixture of ordered and less ordered local structures~\cite{tanaka1998simple,tanaka2000simple,tanaka2012bond}. 
 
Upon cooling liquid water, the experimentally measured FSDP does indeed split into two distinct maxima~\cite{Skinner-14-12, Esmaeildoost-21-12, Sellberg-14-6}. Consistent with Shi and Tanaka's interpretation, this effect reveals the separation between two populations: the low-$q$ peak is associated with locally tetrahedrally coordinated, ordered environments; the high-$q$ peak with more compact, disordered local domains~\cite{Skinner-14-12, Esmaeildoost-21-12, Sellberg-14-6,shi2019distinct,shi2020direct}. The separation between these peaks increases upon cooling, indicating the emergence of spatial correlations and structural differentiation. These features are aligned with water’s known thermodynamic anomalies and the hypothesised liquid–liquid critical point (LLCP) \cite{Poole-92-11}.

Despite these insights, the topological origins of the FSDP splitting in liquid water remain incompletely understood. In particular, the role of medium-range topological motifs, such as hydrogen-bonded rings of various sizes, in shaping the structure factor has not been investigated. Recent work has emphasised the importance of topology in other disordered systems: Zhou \textit{et al.}~\cite{Zhou-21} and Tavanti \textit{et al.}~\cite{Tavanti-20-9} demonstrated that shifts in ring distributions correlate directly with changes in diffraction features in silicate and chalcogenide glasses.

\begin{figure*}[t]
    \centering
    \includegraphics[width=\linewidth]{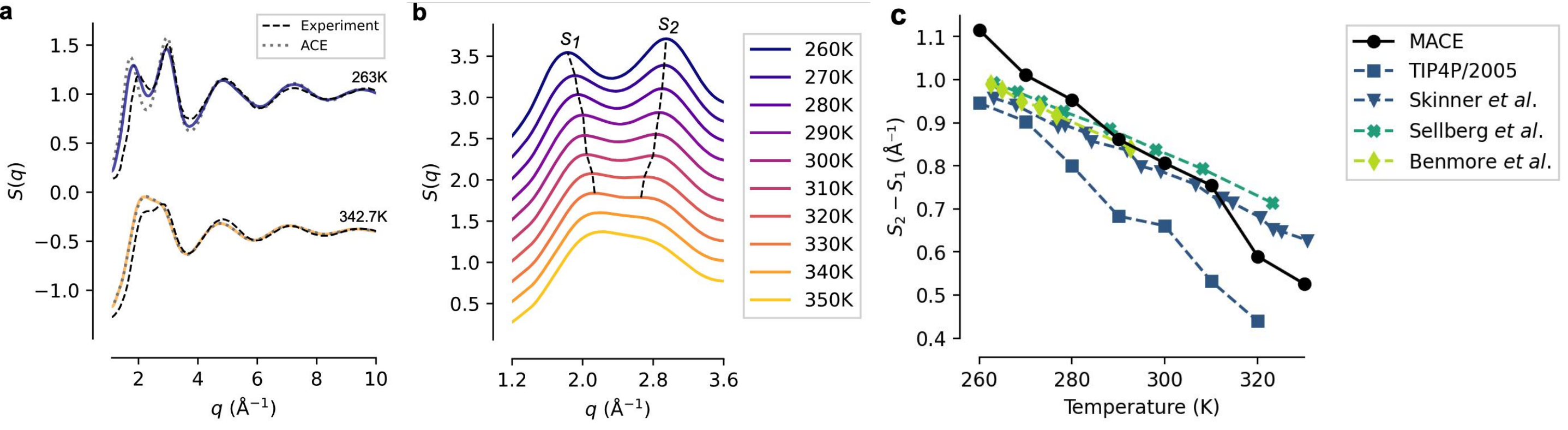}
    \caption{Peak splitting in the structure factor, $S(q)$, of liquid water at 1\,bar. (a) Simulation results obtained with MACE (solid lines) and ACE (dotted lines) potentials at 240\,K (yellow) and 260\,K (dark blue), alongside experimental data (dashed lines) from Ref.~\cite{Skinner-14-12}. The corresponding experimental temperatures are indicated to the right of each curve. (b) Magnified view of the low-$q$ (1.2\,Å$^{-1}$ to 3.6\,Å$^{-1}$) region, highlighting the emergence and divergence of two peaks, $S_1$ and $S_2$, as shown by black dashed lines. Peak positions at each temperature are determined via Gaussian Process (GP) fits to the simulation data. At temperatures above 330\,K, the GP fit does not resolve two separate maxima. (c) Temperature dependence of the peak separation $S_2 - S_1$ comparing the current MACE model (black circles) with the classical TIP4P/2005 model (squares), and experimental data from Skinner \textit{et al.}~\cite{Skinner-14-12} (triangles), Sellberg \textit{et al.}~\cite{Sellberg-14-6} (squares), and Benmore \textit{et al.}~\cite{Benmore-19-9} (diamonds).}
    \label{fig:struct-fact-per-temp}
\end{figure*}

Here, we employ advanced atomistic simulations to investigate precisely this question: how hydrogen-bonded rings of various sizes contribute to $S(q)$. By directly linking specific topological motifs to the observed bifurcation in the FSDP, our findings demonstrate that the topology of the hydrogen-bond network (HBN) plays a central role in shaping water’s structural signatures, and, by extension, its physical properties. This result establishes a clear connection between topological properties and experimental evidence, thus enhancing our understanding of water’s complex nature.

{\em Methodology}---Our main results are obtained using a machine-learned interatomic potential (MLIP) model that we fitted to the dataset of Ref.~\citenum{Ibrahim-24-12} using the MACE architecture~\cite{Batatia-23}. 
Molecular-dynamics (MD) simulations using this model capture the experimentally observed splitting of the FSDP into two distinct maxima (Fig.~\ref{fig:struct-fact-per-temp}a), as well as the corresponding linear increase in their separation at lower $T$~\cite{Skinner-14-12,Esmaeildoost-21-12,Sellberg-14-6} (Fig.~\ref{fig:struct-fact-per-temp}b). We compare our model's predictions to experimental data from Ref.~\citenum{Skinner-14-12}, to simulations with the atomic cluster expansion (ACE) MLIP model from Ref.~\citenum{Ibrahim-24-12}, whose training dataset we have used here, and to the popular classical TIP4P/2005 model of water~\cite{abascal2005general}, which reproduces well many properties of liquid water~\cite{abascal2007dipole,yagasaki2018phase}, including density and compressibility~\cite{Martelli-19-3}, as well
as the phase behaviour of ices~\cite{conde2013determining,wong2015pressure,rescigno2025observation}. 
Overall, our MACE model reproduces the experimental $S(q)$ well, with some discrepancies at low $q$, consistent with differences observed in other validation metrics (Supplementary Materials): it tends to over-structure the network, possibly correlated with the predominance of ice-like configurations in the training dataset~\cite{Ibrahim-24-12}. At high $T$, both MLIPs yield virtually indistinguishable predictions; at low $T$, the MACE model agrees better with experiment than the ACE one, a trend also reflected in the density isobar predictions (Fig.~S2).

A magnified view of the low-$q$ region in Fig.~\ref{fig:struct-fact-per-temp}b highlights the increasing separation between these peaks, S$_1$ and S$_2$, below 330\,K. The peak positions, identified using a Gaussian Process (GP) fit, show a clear divergence as temperature decreases. This trend is evident from the distance between peaks S$_2$ and S$_1$, which we plot in Fig.~\ref{fig:struct-fact-per-temp}c. Our MACE model reproduces the experimentally observed linear increase in separation~\cite{Skinner-14-12, Sellberg-14-6, Benmore-19-9}, while TIP4P/2005 simulations slightly underestimate it.
Above 330\,K, thermal fluctuations obscure the distinct local environments that cause the splitting, and the GP fit resolves only a single broad maximum, corresponding to an averaged O–O separation.

{\em Ring Topology in Liquid Water}---Having established that our model reproduces the key experimental observation, we proceed to analyse the network topology of the liquid based on the MD trajectories. 
\begin{figure*}
    \centering
    \includegraphics[width=.99\linewidth]{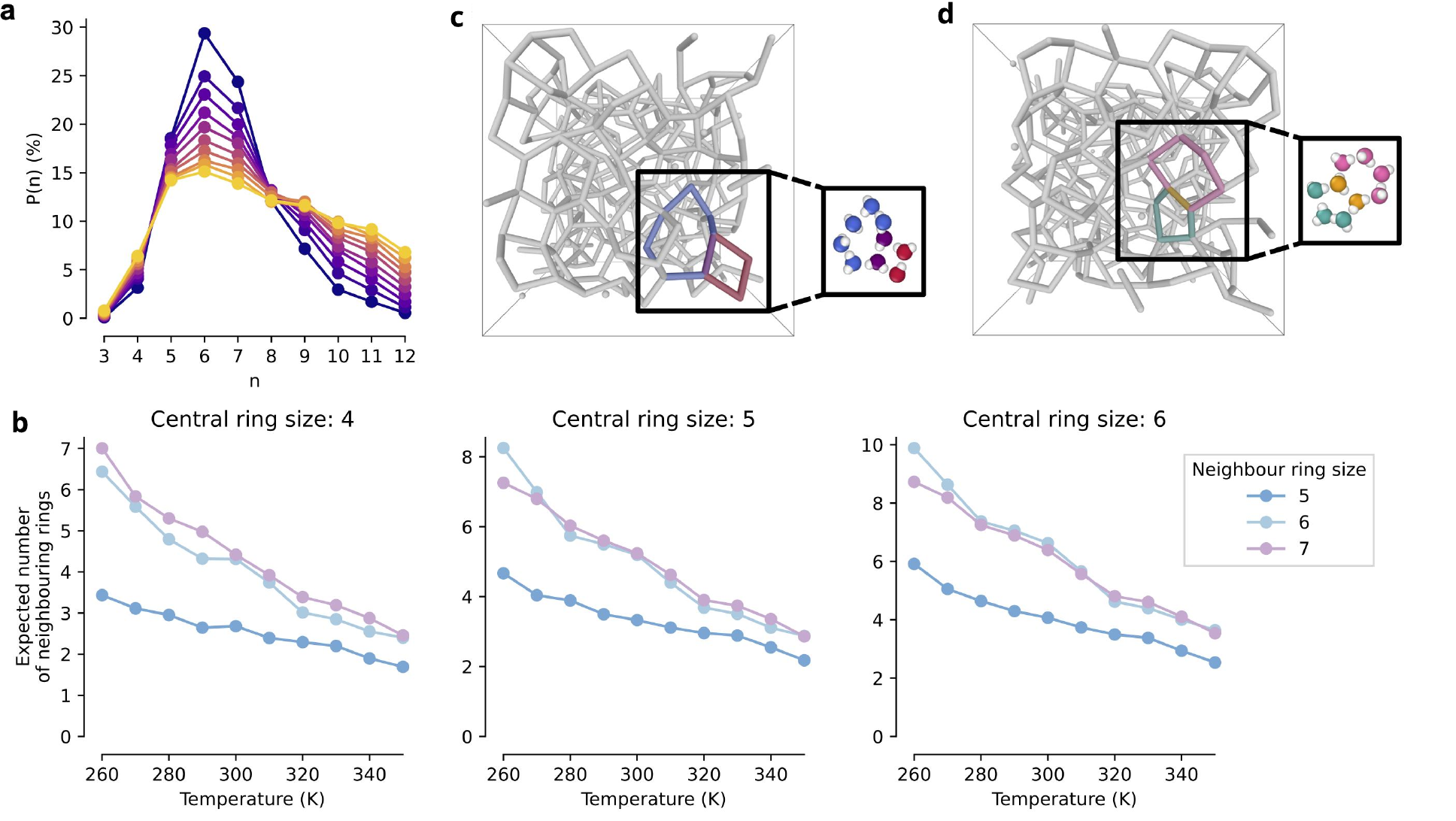}
    \caption{Ring statistics and neighbourhood structure in the HBN of liquid water from 260 to 350\,K. (a) Probability distribution, $P(n)$, of observing hydrogen-bonded rings consisting of $n$ water molecules, where $n \in [3, 12]$, across a range of temperatures. (b) Expected number of neighboring rings of sizes 5 (blue), 6 (grey), and 7 (pink), conditional on a central ring of size 4 (left), 5 (middle), or 6 (right) as a function of temperature. (c--d) Representative snapshots of the HBN at 300\,K, illustrating neighbouring ring structures, hydrogens have been omitted for clarity. Panel (c) shows an example of adjacent 4- and 7-membered rings; panel (d) shows an example of adjacent 5- and 6-membered rings. Insets show full molecular representations of the rings. 
    }
    \label{fig:ring-static-stats}
\end{figure*}
Figure \ref{fig:ring-static-stats}a shows the ring-size distribution in liquid water as a function of temperature. At lower $T$, the distribution shows a clear preference for medium-sized, 5–7-membered rings. At higher $T$, the distribution broadens, reflecting increased thermal fluctuations that disrupt the network and promote the formation of larger rings. 
6-membered rings remain the most probable throughout, highlighting a local topological similarity to cubic and hexagonal ice, the crystalline ground state of water at ambient pressure. Additional analysis, shown in Fig.~S7, confirms that cooling substantially increases the {\em absolute} number of 5--7-membered rings---consistent with the progressive establishment of an extended HBN, as also observed in liquid water described with the TIP4P/2005 model~\cite{Martelli-19-3,martelli2021topology,martelli2022steady} and with other classical interaction potentials~\cite{matsumoto2007topological}.

Next, we examined the topology surrounding rings of specific sizes. Figure~\ref{fig:ring-static-stats}b shows the temperature dependence of the expected number of neighbouring rings of sizes 5, 6, and 7, conditional on a central ring of size 4, 5, or 6 water molecules. Results for rings of other sizes are given in Fig.~S9. Most central rings (from pentagon to hendecagon) prefer 6-membered neighbours, particularly at lower $T$. This preference aligns with water's tendency towards tetrahedral coordination, a structural arrangement best accommodated by hexagonal motifs within the HBN. The similarity in the shape of these neighbouring ring distributions across various central-ring sizes underscores the pervasive influence of the underlying tetrahedral geometry, indicating a consistent local structural motif that propagates throughout the liquid, irrespective of the specific central ring.

A notable exception is observed for tetragonal rings, which are preferentially found next to 7-membered rather than 6-membered rings, an inversion of the general trend observed for most other ring sizes. Figures \ref{fig:ring-static-stats}c and \ref{fig:ring-static-stats}d show two representative snapshots highlighting the connection of a tetragonal with a 7-membered ring, and of a 5-membered with a 6-membered ring, respectively. This non-trivial embedding of tetragons within hexagonal and heptagonal surroundings implies a local topological arrangement that is only possible in a hyperbolic space, i.e., a non-Euclidean space with negative curvature~\cite{Coxeter-99}. This unique topological affinity suggests that tetragonal rings may play an active role in stabilising domains rich in heptagonal rings, and vice versa.

\begin{figure*}
    \centering
    \includegraphics[width=\linewidth]{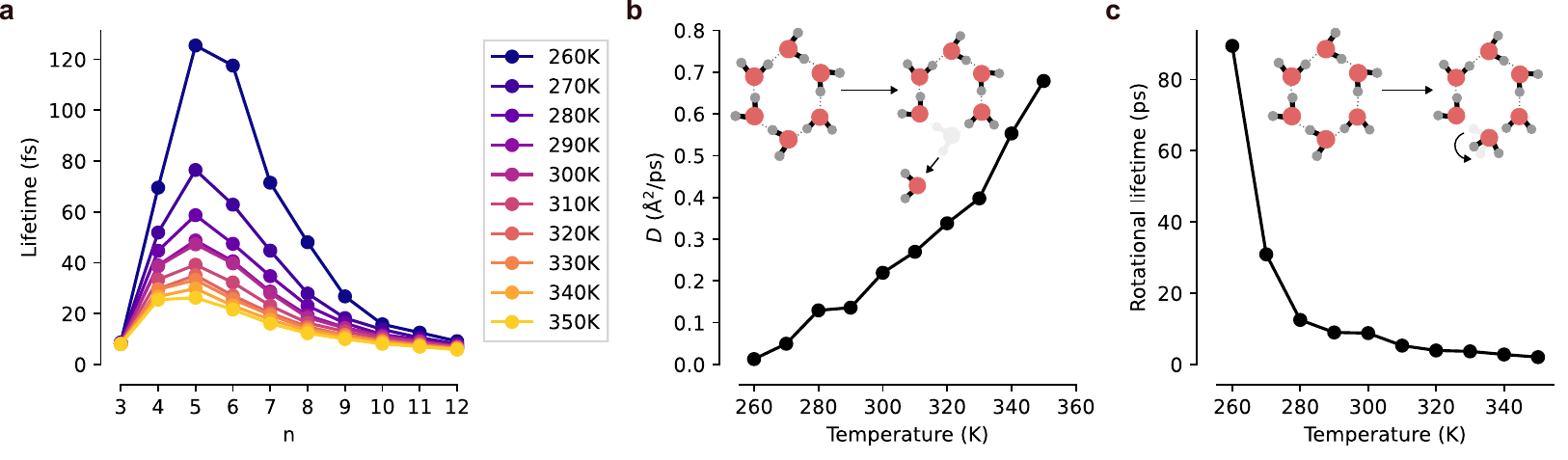}
    \caption{Temperature dependence of dynamic properties of hydrogen-bonded rings in liquid water. (a) Average lifetimes of hydrogen-bonded rings as a function of ring size ($n \in [3, 12]$) and temperature. (b) Self-diffusion coefficient, $D$, of water as a function of temperature. The cartoon illustrates the breaking of a hydrogen-bonded ring via translational motion of a water molecule. (c) Rotational lifetime of water molecules as a function of temperature. The cartoon illustrates the breaking of a hydrogen-bonded ring via rotational motion of a water molecule.}
    \label{fig:ring-dynamic-stats}
\end{figure*}

{\em Lifetimes of Rings}---We now turn to the dynamics of these structural motifs. Figure~\ref{fig:ring-dynamic-stats}a shows the lifetimes of hydrogen-bonded rings as a function of size and temperature. At 260\,K, lifetimes vary substantially depending on ring sizes, ranging from $\sim10$\,fs for smaller or larger rings up to $\sim120$\,fs for hexagonal rings. At this level of supercooling, the lifetime of 5-8-membered rings is longer compared to the classical TIP4P/2005 model (Fig.~S9), possibly because the density of MACE is substantially lower than the experimental and the TIP4P/2005 water model. By 330\,K, the lifetimes converge to a narrow range of 10--30\,fs, suggesting that no particular ring topology remains particularly long-lived; we observe no significant difference with TIP4P/2005 (Fig.~S9). Rings form and dissolve rapidly at high $T$, reflecting a highly transient HBN in which no specific topology dominates the liquid’s organisation, mirrored by the broadening of the ring-size distribution (Fig.~\ref{fig:ring-static-stats}a). This highly dynamic HBN is a requirement for the simulated structure factor to agree with the experimental one~\cite{head2006tetrahedral}.
And yet, a clear lifetime hierarchy persists: pentagonal rings exhibit the greatest lifetime, followed by hexagonal and tetragonal motifs, regardless of $T$. 
Despite their comparable energetic stability, the extended lifetime of pentagonal motifs relative to hexagonal ones can be partially attributed to statistical effects. Specifically, forming a 5-membered ring requires fewer molecules than a 6-membered one, reducing potential disruption points. With fewer participating molecules, the probability of any individual translation or rotation breaking the ring is lower, making these smaller rings statistically less prone to disintegration.

The lifetimes of both 5- and 6-membered rings increase rapidly upon cooling (Fig.~\ref{fig:ring-dynamic-stats}a), highlighting their growing significance in the supercooled regime as the dominant long-lived units. The distinction between both ring types is critical: 6-membered ones promote local order and are associated with slowed molecular dynamics, while 5-membered ones introduce geometric frustration that impedes crystallisation. The near equivalence of their lifetimes at 260\,K suggests a delicate balance between order-promoting and disorder-promoting effects. This interplay may underlie some anomalous behaviours observed in supercooled water, pointing to a possible topological origin of water's dynamic and thermodynamic anomalies~\cite{zimon2025predicting}.

Interestingly, tetragonal motifs exhibit lifetimes comparable to, or even exceed, those of hexagonal rings at high temperatures down to 270\,K, despite being less favourable energetically. Moreover, as shown in Fig.~S8, the temperature dependence of tetragonal ring lifetimes is markedly weaker than that observed for pentagonal and hexagonal rings, indicating that their stability is less affected by thermal fluctuations---thus suggesting the presence of additional stabilising factors, such as local geometric constraints or strain-induced rigidity, that may inhibit the dissolution or rearrangement of tetragonal rings as $T$ increases. Their topological environment can partially explain this unusual kinetic behaviour (Fig.~\ref{fig:ring-static-stats}d): their reconfiguration may be suppressed by being embedded within stabilising hexagonal and heptagonal surroundings. Such cooperative embedding between tetragonal and heptagonal rings could reinforce both motifs, further enhancing the longevity of tetragonal rings.

The dynamic trends observed in Fig.~\ref{fig:ring-dynamic-stats}a are reflected in the macroscopic transport behaviours of the liquid. The self-diffusion coefficient, $D$, increases by a factor of four from 260 to 330\,K (Fig.~\ref{fig:ring-dynamic-stats}b). Simultaneously, the rotational lifetime (Fig.~\ref{fig:ring-dynamic-stats}c) drops steeply from over 80\,ps at 260\,K to below 10\,ps by 330\,K, reflecting faster reorientation, and further supporting the picture of a highly dynamic, transient network at elevated temperatures.

{\em Signatures in the Structure Factor}---To investigate how these persistent topological motifs manifest in experimentally measurable observables, we examine their contributions to the static structure factor. In particular, we ask whether the characteristic peak splitting at low $T$ can be traced back to specific ring sizes, and whether their persistence influences structural signatures of heterogeneity in the liquid. We employ the methodology introduced by Zhou \textit{et al.}~\cite{Zhou-21} to isolate the contributions of individual ring sizes to the structure factor. Specifically, for each ring size, $n$, we identify and retain only the atoms forming $n$-membered rings, remove all other atoms from the simulated structure, and compute the corresponding structure factor $S_n(q)$ for the reduced system~\footnote{It is important to note that this ring-wise decomposition removes all cross-terms between atoms belonging to different rings: hence, the absolute intensities of the decomposed structure factors, $S_n(q)$, are not physically meaningful and do not add up to the total $S(q)$. Instead, the relevant metric in this analysis is the position of the peaks in each $S_n(q)$, reflecting the dominant intermolecular length scales associated with each ring size. The emergence and persistence of peak splitting, in particular, can indicate the structural role of specific ring topologies.}.
\begin{figure*}[t]
    \centering
    \includegraphics[width=\linewidth]{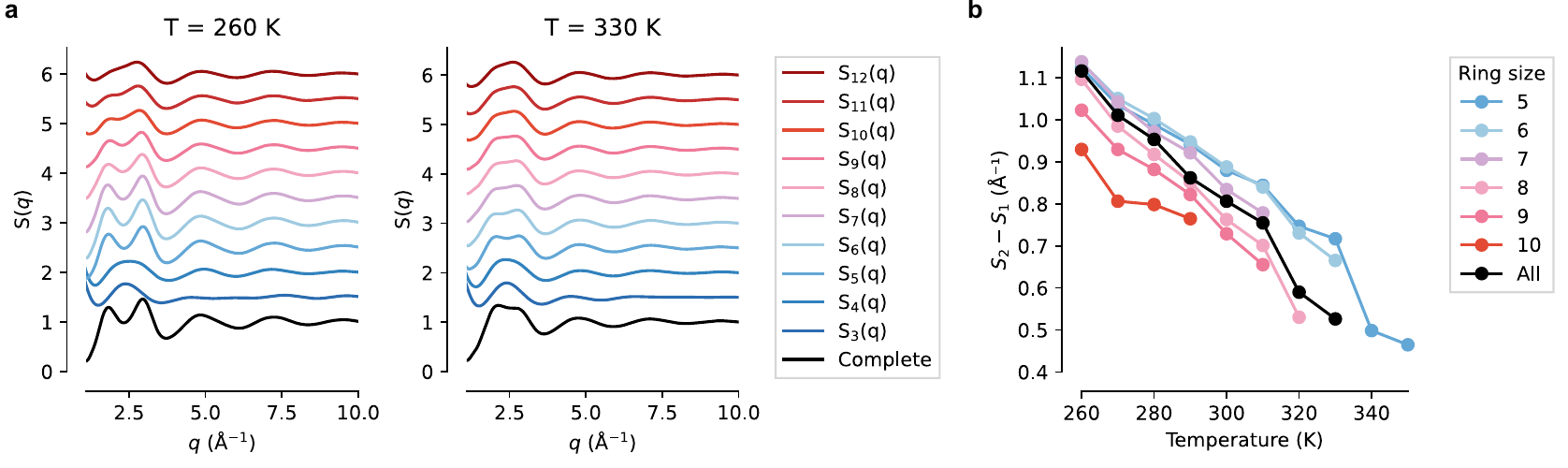}
    \caption{Ring-resolved decomposition of the structure factor, $S(q)$, of liquid water at 1\,bar. (a) Structure factors at 260\,K and 330\,K, decomposed into contributions from individual ring sizes, $S_n(q)$, where $n \in [3, 12]$. The complete structure factor is shown in black, while colored lines represent contributions from specific ring sizes (blue for smaller rings, red for larger rings). (b) Temperature dependence of the peak separation, $S_2 - S_1$, for different ring sizes. The black curve represents the trend for the complete structure factor as shown in Fig.~\ref{fig:struct-fact-per-temp}c.}
    \label{fig:decomp-per-temp}
\end{figure*}
Figure~\ref{fig:decomp-per-temp}a presents this decomposition at the two temperature extremes, viz.\ 260 and 330\,K, where the total $S(q)$ still shows resolvable peak splitting. Additional data for intermediate temperatures are provided in Figure~S10.

At 260\,K, 5--7-membered rings exhibit structure factors most similar to the total $S(q)$, consistent with their prevalence at that temperature (Fig.~\ref{fig:ring-static-stats}a). This prominence also correlates with the relatively long lifetimes of these rings (Fig.~\ref{fig:ring-dynamic-stats}a), which enable them to persist and make a meaningful contribution to the $S(q)$. Their overall abundance (Fig.~S7) reinforces their key roles in shaping the local organisation of the liquid.
Larger rings (octagonal and nonagonal ones) also show bimodality in $S_n(q)$, suggesting their involvement in the structural heterogeneity that drives the peak splitting. In contrast, smaller rings (triangular and tetragonal motifs) show little resemblance to the full $S(q)$, implying a limited role in determining broader structural features. Decagonal, hendecagonal, and dodecagonal motifs exhibit a prominent shoulder near the high-$q$ peak $S_2$, suggesting their association with denser, interstitial-rich structures.

At 330\,K, individual ring contributions to $S(q)$ mainly become indistinguishable from one another and from the total structure factor. This loss of topological selectivity reflects the broader and more uniform ring-size distribution observed at high temperatures (Fig.~\ref{fig:ring-static-stats}a), along with the pronounced reduction in ring lifetimes (Fig.~\ref{fig:ring-dynamic-stats}b) and the corresponding increase in translational and rotational mobility (Fig.~\ref{fig:ring-dynamic-stats}b--c). Since ring structures can be disrupted by both translational and rotational motion, the combination of accelerated dynamics in both modes amplifies the breakdown of persistent ring topologies. Consequently, the $S_n(q)$ become increasingly featureless, echoing the breakdown of well-defined local motifs and the emergence of topological disorder.

Finally, in Fig.~\ref{fig:decomp-per-temp}b, we present the distances between the two resolved peaks for each ring-specific structure factor, $S_{n}(q)$. This analysis highlights that rings sized 5--10 predominantly drive the divergence in peak positions: these most clearly reflect the coexistence of low- and high-density local environments in liquid water.
In particular, we observe that only rings sized 5--10 exhibit peak splittings significant enough to be discerned as two distinct maxima. Notably, pentagonal rings display the most persistent peak splitting, with both $S_1$ and $S_2$ being resolved up to 350 K. In contrast, the total structure factor distinguishes two peaks only up to 330 K.

The persistence of peak splitting in pentagonal rings up to higher temperatures suggests that these structures maintain distinct local ordering even as the overall network becomes more disordered. This behaviour underscores the significance of pentagonal rings in influencing water's anomalous properties, particularly in the supercooled regime where the balance between order-promoting and disorder-promoting motifs becomes critical.
Conversely, as the ring size increases, the separation between the length scales of locally ordered and locally disordered structures becomes progressively obscured, resulting in convergence into a single resolved peak at lower temperatures (red lines in Fig.~\ref{fig:decomp-per-temp}).

{\em Conclusions}---We have linked the ring topology in liquid water to the experimentally observed splitting of its FSDP.
Extending beyond previous interpretations~\cite{Skinner-14-12, Esmaeildoost-21-12, Sellberg-14-6,shi2019distinct,shi2020direct}, our results demonstrate that the observed peak splitting reflects specific medium-range topological motifs. In particular, pentagonal rings persistently contribute to this feature, highlighting their unique role in sustaining structural heterogeneity via geometric frustration. This supports the idea of a subtle competition between order (hexagonal rings) and disorder (pentagonal rings) in supercooled water~\cite{Martelli-19-3}. Additionally, the unexpected stability of tetragonal rings—especially their tendency to cluster near heptagonal ones—points to a complex network organisation. These correlations suggest that the local environment plays a stabilising role even for topologically strained motifs.
Future research could extend this ring-based analysis to other liquid and amorphous materials, and further to atomistic simulations of biological systems, which are becoming increasingly accurate~\cite{omranpour2024perspective}: water plays a crucial role in protein folding, enzyme activity, and membrane dynamics, and topological analyses could provide new insights into how the local environment of water molecules influences longer-range structure and function.

\begin{acknowledgements}
Z.F.B. was supported through an Engineering and Physical Sciences Research Council DTP award EP/X524979/1 and IBM Research.
V.L.D. acknowledges support from the John Fell OUP Research Fund.
F.M. is grateful to Francesco Sciortino and Pablo G. Debenedetti for useful discussions. Data supporting this work are available at \href{https://doi.org/10.5281/zenodo.18187274}{Zenodo (DOI: 10.5281/zenodo.18187274)}, \href{github}{https://github.com/ZoeFaureBeaulieu/water-topological-analysis/tree/main}.
\end{acknowledgements}

\bibliography{pnas-sample}

\end{document}